\documentclass[referee,usegraphicx,usenatbib]{mn2e}

\title[Observations of three slow glitches in PSR B1822$-$09]
      {Observations of three slow glitches in the spin rate of
       the pulsar B1822$-$09}
\author[T. V. Shabanova]
       {T. V. Shabanova\thanks{E-mail:tvsh@prao.psn.ru}
        \\
        Astro Space Center, P.N. Lebedev Physical Institute,
        Leninskij Prospect 53, 117924 Moscow, Russia}
\begin{document}

\date{2004 November 02}

\pagerange{\pageref{firstpage}--\pageref{lastpage}} \pubyear{2004}

\maketitle

\label{firstpage}

\begin{abstract}

Three slow glitches in the rotation rate of the pulsar B1822$-$09
were revealed over the 1995--2004 interval. The slow glitches observed
are characterized by a gradual increase in the rotation frequency with
a long timescale of several months, accompanied by a rapid decrease in
the magnitude of the frequency first derivative by $\sim$ 1--2 per cent
of the initial value and subsequent exponential increase back to its
initial value on the same timescale. The cumulative fractional increase
in the pulsar rotation rate for the three glitches amounts to
$\Delta\nu/\nu_{0} \sim 7 \times 10^{-8}$.

\end{abstract}


\section{Introduction}

Many pulsars exhibit significant variations in their rotation
rates superimposed on the deterministic pulsar spin-down due to
electromagnetic torque. Rotation variations may occur in the form
of glitches - sudden increases in rotation frequency and in the
form of timing noise - random variations in pulse arrival times.
For slow pulsars, a rotation phase is well described by a simple
${\nu},\dot{\nu}$ spin down model. The values of the second
derivative $\ddot\nu$ due to slowdown are usually very small in
comparison to the measurement uncertainties. The significant
second derivatives measured for most slow pulsars are manifestations
of random walk processes that produce timing residuals with cubic
polynomial components \citep{co2}. These cubic structures make up
30 per cent of many different structures observed in the timing
residuals \citep{hob}. The strength of timing noise
is correlated with the period derivative and is quantified by
the stability parameter based on a non-zero second derivative of
the rotation frequency \citep{co1,arz}.

PSR B1822$-$09 is of great interest as it enables the timing behaviour
of a pulsar to be studied over a long observing span.
It has a period of 0.769 s,
a large period derivative of $52.36 \times 10^{-15}$, implying
a relatively young characteristic age of $\sim$ 230 kyr. An analysis
of timing data over the observational interval 1989--1993 by
\citet{arz} showed that this pulsar possesses considerable timing
noise described by the stability parameter $\Delta_{8}$=-1.2.
Timing observations for a longer interval of 3000 days \citep{lyn}
before the glitches occurred also showed that this pulsar is among
the noisiest pulsars. During this interval, variations in pulse
arrival times for the pulsar were characterized by red power spectrum
and consequently, timing residuals relative to a simple spin-down
model were described by a cubic term corresponding to
the significant second derivative.

In 1994 September the pulsar suffered a small glitch with
${\Delta\nu}/{\nu} \sim 0.2 \times 10^{-9}$ \citep{sh1}. This
glitch was typical, associated with a sudden increase in the
rotation frequency. Subsequent monitoring of the pulsar showed
evidence for a new kind of timing variations which may be
described in terms of a slow glitch. Since 1995, the pulsar has
suffered three slow glitches. The present paper is the third in
series studying the glitch behaviour of the pulsar B1822$-$09. The
signature of the first slow glitch occurred in late 1995 was
described in the first paper \citep{sh1}. It was shown that the
slow glitch observed was characterized by a gradual increase in
the rotation frequency during 620 days, accompanied by a rapid
decrease in the magnitude of the frequency first derivative by
$\sim$ 0.4 per sent and subsequent increase back to its initial
value on the same time span. In the second paper \citep{sh2}, the
glitch behaviour of the pulsar for the period 1991--1998 was
studied at widely separated frequencies of 0.1 and 1.6/2.3 GHz
using quasi-simultaneous observations made at the Pushchino Radio
Astronomy Observatory (PRAO) and the Hartebeesthoek Radio
Astronomy Observatory (HartRAO). The authors showed that the $\nu$
and $\dot\nu$ changes with time are similar at both observational
frequencies and reported the second large decrease in the
magnitude of the frequency derivative by $\sim$ 2.4 per sent which
occurred in 1998 August. The present paper summarises the results
of the two previous papers, reports the third slow glitch occurred
in 2000 December and presents the description of the timing
behaviour of PSR B1822$-$09 over the 19-yr data span from 1985 to
2004.

The third slow glitch was independently detected by \citet{zou}
and published in a recent paper obtained from astro-ph after submission
of the present paper. Comparison between the results presented in
both papers is given in the results section.

\section[]{Observations and timing analysis}

Timing observations of the pulsar have been made at the Pushchino
Observatory at frequencies around 102.5 and 112 MHz with the BSA
transit radiotelescope, using a 32 $\times$ 20 kHz filter bank
receiver. Observations have been conducted since 1991 a few times
per month and lately daily. The detected signal was sampled
at 1.28-ms intervals and integrated synchronously with the apparent
pulsar period for 3.2 minutes of the BSA transit time.
After the dispersion removal the signals from all the channels
were summed to form an average pulse profile for a single observation.

The topocentric arrival times for each observation were calculated
by cross-correlating an average pulse profile with a low-noise
template and then were corrected to the barycenter of the Solar
System using the TEMPO software package\footnote
{http://www.atnf.csiro.au/research/pulsar/timing/tempo}
and the JPL DE200 ephemeris. A pulsar position and a proper motion
assumed in this correction were taken from \citet{arz}
and \citet{fom}, respectively.
A simple spin-down model involving a rotation frequency $\nu$
and its first derivative ${\dot\nu}$ was used for fitting
to the barycentric arrival times giving the pulse phase $\phi$
at the time $t$ as

\begin{equation}
\phi(t) = \phi_{0} + \nu(t-t_{0}) + {\dot\nu}(t-t_{0})^{2}/2 ,
\end{equation}

where $\phi_{0}$ is the phase at some reference time $t_{0}$.
The differences between the observed times and times predicted
from a best fit model gave the timing residuals which were used
for an analysis of the pulsar rotation behaviour.

\section[]{Results}

In order to study variations in the spin-down parameters of the pulsar
in more detail, the rotation frequency and frequency first derivative
were calculated by performing local fits to the arrival time data over
the intervals of 120--200 days. The timing data set analysed includes
the Pushchino data collected for the period 1991--2004 and the HartRAO
data collected over the 1985--1998 interval and taken from the previously
published paper \citep{sh2}.

\subsection{Signature of the first slow glitch}

\begin{figure}
 \centering
 \includegraphics[width=120mm]{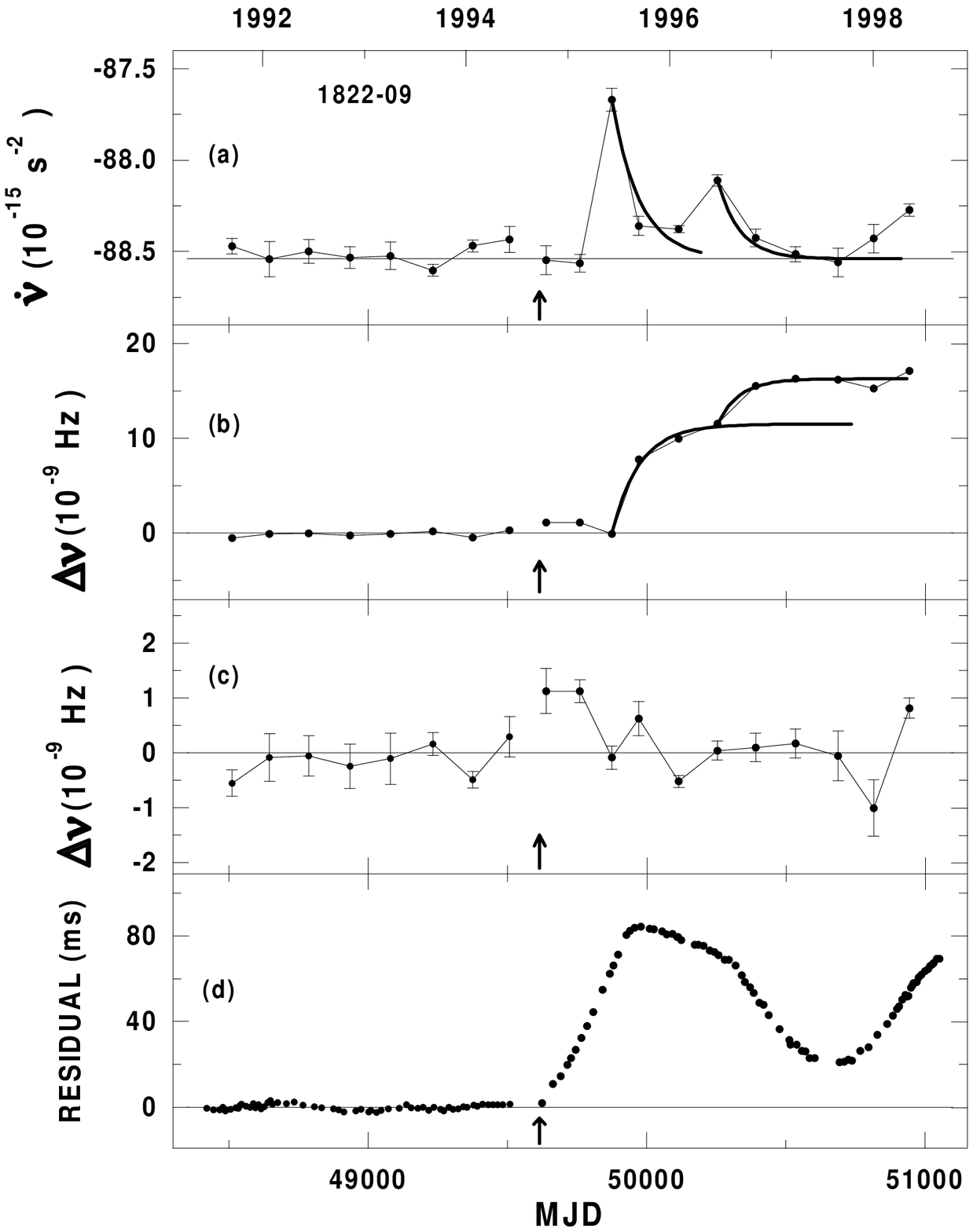}
 \caption{Signature of a typical glitch and slow glitch observed
          in PSR B1822$-$09.
  a) $\dot\nu$ versus time for a typical glitch occurred in September
  1994 and a slow glitch occurred in August 1995. The horizontal line
  denotes the mean value of $\dot\nu_{0}$ over the 1991--1994 interval.
  b) $\Delta\nu$ versus time show the signature of the 1994 glitch
  and the slow glitch.
  c) As for (b) but with the residuals for the slow glitch after
  subtracting the two asymptotic exponentials on timescales
  of 100 and 80 days.
  d) Timing residuals relative to independent fits for $\nu$ and
  $\dot\nu$ to the data $\sim$ 1200 days before the 1994 glitch and
  1400 days after the 1994 glitch.
  Arrows indicate the time at which the 1994 glitch occurred.
  The exponential fits drawn as thick-lines in panels (a) and (b).
   }
\label{glitch}
\end{figure}

This section presents a detailed description of the signature
of the first slow glitch, using the $\nu$,$\dot\nu$ data
published in the two previous papers \citep{sh1,sh2}.
During the period 1991--1998, the pulsar suffered two glitches,
the first of which occurred at the end of September, 1994 (MJD 49615)
and had a typical signature, while the second occurred about a year
later (MJD 49940) and exhibited an unusual signature related to
a gradual increase in the pulsar rotation frequency.
The plots of $\dot\nu$ and $\Delta\nu$ together with the timing
residuals over the interval 1992--1998 are shown in Fig.~\ref{glitch}.
The frequency residuals
are given relative to a simple slow-down model, involving the mean
value of $\nu_{0}$ and $\dot\nu_{0}$ defined over the interval
1991--1994, where the behaviour of the pulsar was steady. The
values of $\dot\nu$ and $\Delta\nu$ were calculated from the local
fits, performed over intervals of 120 days. The plotted points
agree well with the mean spin-down parameters up to the 1994
glitch. This glitch was quite small with the fractional increase
of rotational frequency equal to $0.8 \times 10^{-9}$. Apparently,
the glitch was followed by a relaxation in frequency as shown in
Fig.~\ref{glitch}b.

From Figs.~\ref{glitch}a and~\ref{glitch}b, it is seen that
$\dot\nu$ and $\Delta\nu$ begin to vary considerably shortly after
the 1994 glitch. During the $\sim$ 500-day interval from the
middle of 1995 to the end of 1996, $\dot\nu$ was smaller than the
mean spin-down. The observed changes in $\dot\nu$ involve two similar
events. The first one occurred at about MJD 49874 when $\dot\nu$
rapidly decreased by approximately 0.7 per cent of the initial value,
while the second one occurred at about MJD 50253 when $\dot\nu$
decreased by $\sim$ 0.3 per cent of the initial value. Both
the events were followed by an increase back to its initial value.
The relaxations are well described by two different exponential
components with timescales of 100 and 80 days, respectively, as
shown in Fig.~\ref{glitch}a.

Fig.~\ref{glitch}b shows that a slow increase of the frequency
residuals relative to the pre-glitch data begins when $\dot\nu$
rapidly reaches its minimum value (near MJD 49874).
The slow growth of $\Delta{\nu}$ is well described by a combination
of two asymptotic exponential functions with the same timescales
of 100 and 80 days. The frequency residuals after subtracting these
asymptotic exponentials are given in Fig.~\ref{glitch}c.
The size of the slow glitch after a span of two years is equal to
${\Delta\nu}/{\nu_{0}} = 13 \times 10^{-9}$. Apparently, the slow
glitch observed is about one order of magnitude larger than
the 1994 glitch which showed a typical signature.

The timing residuals relative to the best-fit for $\nu$ and $\dot\nu$
for the data from the interval of 1200 days preceding the 1994 glitch
and for the data from the interval of 1400 days following the 1994
glitch are presented in Fig.~\ref{glitch}d.
The timing residuals for the data over the period 1995--1998 when
the slow glitch was observed, have quasi-periodic structure with
the amplitude of about 80 ms.
From Figs.~1b and~1d, it is clear that the changes in the slope of
the curves of frequency and timing residuals reflect the detailed
structure in the frequency first derivative $\dot\nu$.

\subsection{Observed properties of the slow glitches}

\begin{figure}
 \centering
 \includegraphics[width=120mm]{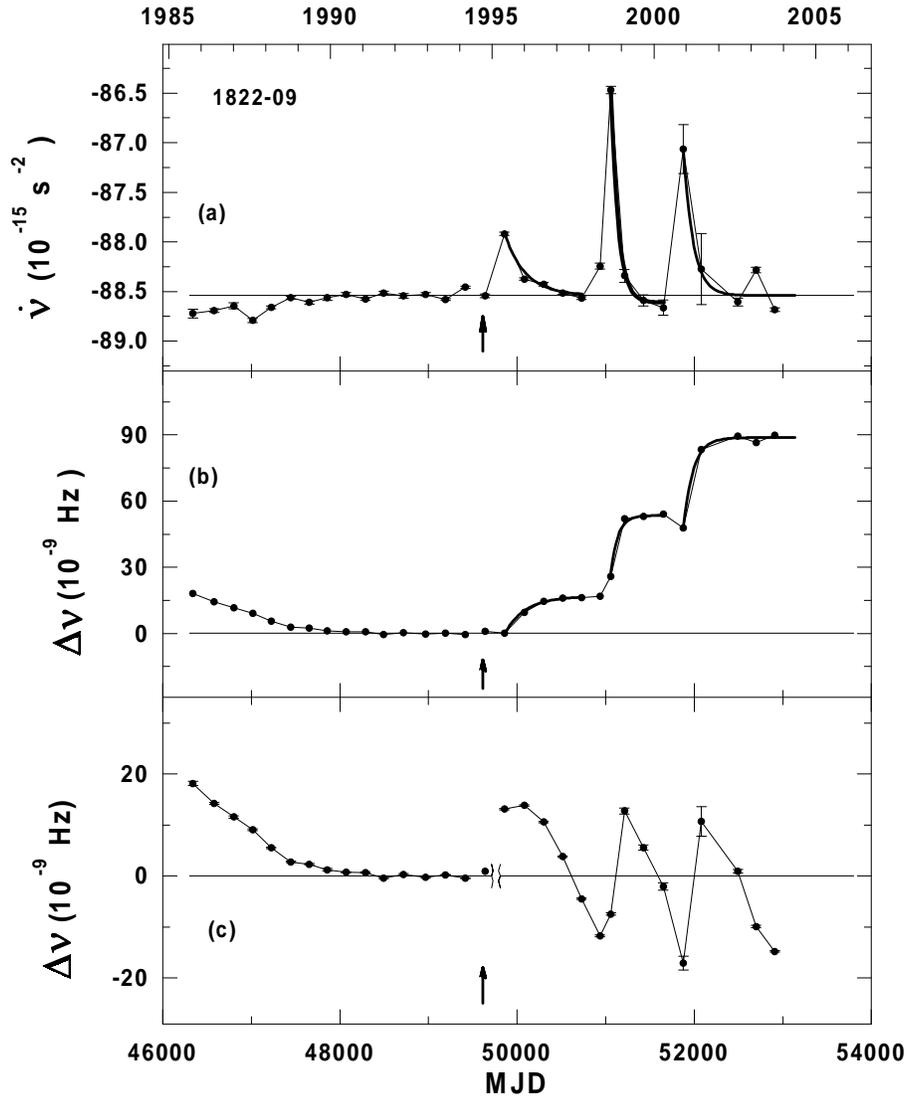}
 \caption{The frequency first derivative and frequency residuals
  for PSR B1822$-$09 from 1985 to 2004.  The 1994 glitch and three
  slow glitches during the 1995--2004 interval are clearly seen
  in the data.
  a) $\dot\nu$ versus time. Three rapid decreases in $\dot\nu$
  over the interval 1995--2004 are the effect of the slow glitches.
  The horizontal line denotes the mean value of $\dot\nu_{0}$ over
  the 1991--1994 interval.
  b) $\Delta\nu$ relative to a fit to the data for the interval
  1991--1994. Significant cumulative change in the rotation rate
  is the effect of the three slow glitches.
  c) As for (b) but with $\Delta\nu$ relative to a new fit to
  the data for the 1995--2004 interval where the slow glitches
  occurred.
  Arrows indicate the time at which the 1994 glitch occurred.
  The exponential fits drawn as thick-lines in panels (a) and (b).
  }
\label{locfit}
\end{figure}

Fig.~\ref{locfit}a shows $\dot\nu$ as a function of time over the
entire data span of 19 years from 1985 to 2004. A plot of
$\Delta\nu$ versus time is given in Fig.~\ref{locfit}b. As before,
the frequency residuals are presented relative to the data from
the interval 1991--1994. Here, the local fits were performed over
intervals of 200 days. The dependencies of $\nu$ and $\dot\nu$ on
time are different for the timing data collected before and after
the 1994 glitch. A long-term linear trend in  $\dot\nu$ is well
seen in the pre-glitch data. Both the plots show that there is a
large pre-glitch frequency second derivative $\ddot{\nu}=8.9
\times 10^{-25} s^{-3}$. It is most likely related to timing
noise.

The behaviour of the three slow glitches after 1995 can be clearly
seen in Figs.~\ref{locfit}a and~\ref{locfit}b.
The signature of the first slow glitch was analysed in detail
in the previous section. Common properties of the slow glitches
observed may be described as follows.
The slow glitches are characterized by a rapid initial decrease in
the magnitude of $\dot\nu$ and subsequent increase back to its
previous value with a long timescale of several months. This
causes a gradual growth in the frequency residuals $\Delta\nu$ with
the same timescale. The epoch of a slow glitch is the time at which
$\dot\nu$ rapidly reaches its minimum value. This epoch is slightly
earlier than the epoch when the timing residuals begin to show
the negative change after the slow glitch.

The derived parameters of the three slow glitches observed in the
rotation rate of the pulsar B1822$-$09 are given in Table~1. A
fractional decrease of the frequency derivative is about
0.7, 2.7 and 1.7 per cent for the three events, respectively. The
form of the subsequent increase back to its initial value
is well modelled by an exponential function with a timescale of
235, 80 and 110 days, respectively. A gradual growth in rotation
frequency $\nu$ is well described by asymptotic exponential
functions with the same timescales of 235, 80 and 110 days. A
fractional increase in frequency of $\Delta\nu/\nu_{0}$ is about
13, 20 and 31 $\times 10^{-9}$ for the three slow glitches,
respectively.

\begin{table}
 \centering
 \begin{minipage}{140mm}
 \label{gltpar}
  \caption{Parameters of the three slow glitches}
  \begin{tabular}{@{}llll@{}}
  \hline
 Glitches                &    1    &    2    &    3      \\
\hline
 Epoch (MJD)             &  49857  &  51060  &  51879    \\
 ${\Delta\dot\nu}/{\dot\nu_{0}}$
                         & -0.007  & -0.027  & -0.017    \\
 ${\Delta\nu}/{\nu_{0}}$
 $\,(10^{-9})$           &   12.6  &   20    &   31.4     \\
 $\tau$ (days)           &   235   &   80    &   110     \\
\hline
\end{tabular}
\end{minipage}
\end{table}

The integrated effect of the three slow glitches at the present time
(for example, on May 25, 2004) is equal to
$\Delta\nu \sim 89 \times 10^{-9}$ Hz that gives the total fractional
increase in the rotation frequency of
$\Delta\nu/\nu_{0} \sim 68 \times 10^{-9}$. This means that
at the indicated epoch the pulsar period is approximately 53 ns less
than the expected value from extrapolation of the pre-glitch
1991--1994 model. As a result, the pulsar begins to rotate faster
than it would if the slow glitches had not occurred.

The third glitch was also detected by \citet{zou} at the observing
frequency of 1540 MHz and described as slow glitch, too. The size
of this glitch with ${\Delta\nu}/{\nu} = 31.2(2) \times 10^{-9}$
is in good agreement with the result in Table~1. However, there is
a discrepancy between the quoted values of the changes in the
frequency derivative. \citet{zou} give ${\Delta\dot\nu}/{\dot\nu}
= 1.9(1)\times 10^{-3}$, that is one order magnitude smaller than
${\Delta\dot\nu}/{\dot\nu}$ given in Table~1. The signature of
${\Delta\nu}$ and $\dot\nu$ shown in their Fig.2 over the interval
MJD 51500--53000 is consistent with that plotted in
Figs.~\ref{locfit}a and~\ref{locfit}b of the present paper. The
small event in ${\dot\nu}$ reported by \citet{zou} as the
next possible slow glitch between MJD 52798 and 52969 is clearly
seen in our plot, too. The good agreement between the results
describing the signature of the third slow glitch at different
frequencies of 112 and 1540 MHz provides strong evidence for the
existence of unusual glitch phenomenon in the pulsar B1822$-$09.

The frequency residuals relative to the new model fit involving
the new values of $\nu$ and $\dot\nu$ from the interval 1995--2004
where the three slow glitches occurred, are presented in
Fig.~\ref{locfit}c on the right hand plot.
It is clearly seen that the rotation frequency of the pulsar relative
to the new fit shows the oscillation behaviour. The timescale of this
oscillation can be estimated to be $\sim$ 1000 days.

\subsection{Timing residuals of the pulsar}

\begin{figure}
 \centering
 \includegraphics[width=120mm]{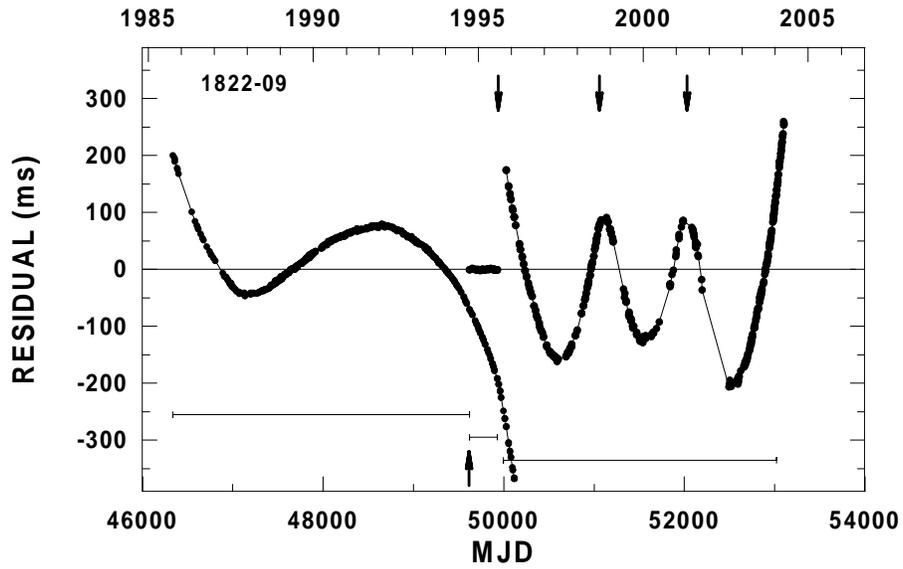}
 \caption{Timing residuals for PSR B1822$-$09 between 1985 and 2004
  relative to a fit to the data for the following three intervals:
  1985--1994, 1994--1995 and 1995--2004.
  The length of each interval is indicated by the horizontal line.
  The bottom arrow indicates the epoch at which the 1994 glitch occurred.
  The three upper arrows indicate the epochs at which the three slow
  glitches occurred.
   }
\label{resid}
\end{figure}

A full picture of the timing residuals for the pulsar B1822$-$09
over the 19-yr data span between 1985 and 2004 is presented in
Fig.~\ref{resid}. Analysis of the entire data set showed that all
arrival times cannot be described by a simple spin-down model
within half the pulsar period because of the presence of several
glitches. Therefore, the timing residuals were obtained from three
independent polynomial fits for $\nu$ and $\dot{\nu}$ over three
different intervals indicated in the plot by the horizontal lines.
The corresponding mean values of the spin-down parameters are
given in Table~2. The errors are given in units of the last quoted
digit. The derived models are a good representation of the timing
behaviour of the pulsar for the total data span of 19 years and
successfully predict the pulse arrival times and the pulse periods
over the present observing session.

For the first interval 1985--1994, the timing residuals relative
to a simple spin-down model show a large cubic term which corresponds
to a large frequency second derivative $\ddot{\nu}=8.9 \times
10^{-25} s^{-3}$. The magnitude of $\ddot{\nu}$ for this pulsar is
a measure of the amount of timing noise \citep{lyn}. It gives the
timing noise parameter $\Delta_{8}$=-0.94 that agrees well with
that of \citet{arz}. The derived braking index, $n\,
=\,{\nu}{\ddot\nu}/{\dot\nu}^{2} \sim 145$, is too large to be due
to the pulsar slowdown and indicates the high level of timing
noise, too. This timing behaviour was interrupted by a small
glitch occurred in 1994 September.

The second interval is a short one-year interval 1994--1995 following
the 1994 glitch and preceding the first slow glitch in 1995 August.
During this interval the pulsar exhibits the timing residuals
dominated by the random deviations at a level of a few milliseconds
with a zero mean.

The third interval is a long interval from 1995 to 2004 where
the three slow glitches occurred. The plot shows that variations
in the pulse arrival times have oscillatory structure with a large
amplitude of about 150 ms. Three cycles of this structure are caused
by the three slow glitches observed for this interval.
It is seen that the timing residuals presented in Fig.~\ref{resid}
well correspond to the frequency residuals of Fig.~\ref{locfit}.
Comparison between the values of the period derivative for
different intervals indicated in Table~2 shows that the presence of
slow glitches decreases significantly the mean value of the period
derivative.

\begin{table}
 \centering
 \begin{minipage}{140mm}
  \label{param}
  \caption{Derived timing parameters of PSR B1822$-$09}
  \begin{tabular}{@{}llll@{}}
  \hline
 Fit interval &   Epoch    &     Period        & Period derivative  \\
     (MJD)    &   (MJD)    &      (s)          &    $(10^{-15})$    \\
 \hline
 46338--49618 & 46338.7577 & 0.76897015431(9)  &    52.3846(5)      \\
 49618--49928 & 49618.8354 & 0.76898499734(6)  &    52.3255(37)     \\
 49996--53034 & 49996.8689 & 0.76898671206(16) &    52.0924(12)     \\
\hline
\end{tabular}
\end{minipage}
\end{table}

\section{Discussion}

The main result of this paper is detailed description of a new
kind of timing variations which may be explained in terms of a
slow glitch. Results of the two previous papers \citep{sh1,sh2}
and also the present paper have shown that the pulsar B1822$-$09
suffered three slow glitches over the 1995--2004 interval. The
third slow glitch was independently detected by \citet{zou} and
the authors have interpreted unusual glitch phenomena in PSR
B1822$-$09 as slow glitches, too. Characteristic properties of the
slow glitches observed are a gradual increase in the rotation
frequency with a long timescale of several months, accompanied by
a rapid decrease in the magnitude of the frequency derivative by
$\sim$ 1--2 per cent of the initial value and subsequent
exponential increase back to its initial value with the same
timescale. The size of a slow glitch after a span of a few years
is rather moderate, with  the fractional increase
$\Delta\nu/{\nu}\sim 20 \times 10^{-9}$. An obvious relaxation in
frequency after a slow glitch is not observed. An analysis of
observations of large samples of pulsars
\citep{co2,gul,co1,lyn2,lyn1,ale,joh,she,wan,kra} showed that an
event similar to a slow glitch was not observed earlier in any
pulsar.

Fig.~\ref{locfit} shows that the small glitch of 1994 lies between
two regions with different behaviour of the frequency derivative.
It is of interest whether this glitch could affect the character of
timing variations and whether this small glitch could act as
a trigger for a slow glitch, about two orders of magnitude larger.

It is known that variations in the pulsar rotation rate in the form
of glitches and timing noise are an important source of information
on neutron star interiors \citep{bay,alp,pin}. These events are common
to many pulsars and arise from sudden and irregular transfer of
angular momentum between a more rapidly rotating interior superfluid
and a solid crust of a neutron star.
Pulsar glitches are characterized by a sudden increase in rotation
frequency with $\Delta\nu/\nu \sim 10^{-10}$ to $10^{-6}$,
followed by a post-glitch relaxation.
They are accompanied by an increase in the magnitude of the frequency
derivative with $\Delta\dot\nu/\dot\nu \sim 10^{-3}$ to $10^{-2}$,
which decays after the glitch \citep{she}. The pulsar glitches and
post-glitch relaxation  reflect changes in the angular momentum
distribution inside a neutron star \citep{alp1}.

The signature of slow glitches, as they have been observed in the
spin rate of the pulsar B1822$-$09, is quite different.
The existence of a gradual glitch, as a response of a
neutron star to a perturbation in its temperature, was predicted
by a model in which a glitch was associated with a thermal
instability in a neutron star \citep{gre}. A slow glitch may be a
thermal response of a neutron star to a sudden local increase of
the inner crust temperature \citep{lin}. However, the behaviour of
the frequency derivative - a rapid initial decrease and subsequent
exponential increase back to the initial value, is not accounted for
by the thermal glitch model. It seems more likely, the significant
variations in spin-down rate have to be attributed to variations
in braking torque. A decrease in spin-down rate requires a
corresponding decrease in torque that brakes rotation of the
pulsar crust.

Torque variations may be caused by changes in the magnetosphere
structure, for example, the variations of the polar cap size. The
region of magnetosphere which rigidly corotates with a neutron
star is restricted to the light cylinder. In order to decrease
$\dot\nu$, the region of a corotating magnetosphere should
decrease, i.e. closed magnetic field lines should temporarily open
and then the polar cap size should increase. The measured
oscillatory behaviour in the rotation frequency $\nu$ on
the timescale of $\sim$ 1000
days reflects the oscillatory changes in torque which suggests the
existence of a long-term oscillation in the polar cap size. This
oscillation could be accompanied by observable changes in an
average pulse profile shape which is determined mainly by the
structure of the magnetic field.

It should be noted that the pulsar B1822$-$09 is a very
interesting object. It exhibits simultaneously four rare
properties of the pulsar emission. Three unusual properties as
mode-changing of the average pulse profile, drifting subpulses and
interpulse emission were revealed by \citet{fow} in the frequency
range 1620--2650 MHz. The forth property - the microstructure in
the main pulse emission, was detected by \citet{gil} at the
frequency of 1420 MHz. At the low frequency of 112 MHz, these
unusual properties are not seen. An analysis of the width and
intensity of the average pulse profile made for the data span
before and after the glitches did not show any changes within the
precision of the measurements. It is reasonable to search for a
correlation between the times of slow glitches and the occurrence
of mode-changing and other properties of the pulsar emission in
the data obtained at high frequencies where the mentioned
properties are clearly seen.

\section*{Acknowledgments}

The author thanks Yu.P. Shitov for useful discussions, the staff
of PRAO for assistance in the observations. The author is grateful
to the referee for helpful comments that have improved
the presentation of the paper.

\newpage
\clearpage

\newpage
\clearpage

\newpage
\clearpage

\bsp

\label{lastpage}

\end{document}